# Wurtzite (Ga,Mn)As nanowire shells with ferromagnetic properties

J. Sadowski,[a,b,c] S. Kret,[c] A. Šiušys,[c] T. Wojciechowski,[c], K. Gas,[c] M. F. Islam,[b] C. M. Canali[b] and M. Sawicki[c]

[a] MAX-IV laboratory, Lund University, P.O. Box 118, SE-221 00 Lund, Sweden.
[b] Department of Physics and Electrical Engineering, Linnaeus University, SE-391 82 Kalmar, Sweden.
[c] Institute of Physics, Polish Academy of Sciences, al. Lotników 32/46, PL-02-668 Warszawa, Poland.

(Ga,Mn)As in wurtzite crystal structure, is coherently grown by molecular beam epitaxy on the {1100} side facets of wurtzte (Ga,In)As nanowires and further encapsulated by (Ga,Al)As and low temperature GaAs. For the first time a true long-range ferromagnetic magnetic order is observed in non-planar (Ga,Mn)As, which is attributed to a more effective hole confinement in the shell containing Mn by a proper selection/choice of both the core and outer shell materials.

## Introduction

Complementary resources of charge and spin degrees of freedom are expected to overcome the inevitable limits that conventional Si-based complementary metal-oxide-semiconductor (CMOS), technology will face in the years to come. Various schemes have been proposed, mainly based on taming of spin currents,[1,2] or on employing radical new architectures.[3] Particularly in the latter context ferromagnetic nanocylinders and nanotubes offer great advantages over their lateral counterparts. Namely, by preventing instable movements of domain walls above a certain critical velocity (the Walker breakdown[4]), extremely high operational speeds and stability of non-volatile magnetic devices are envisaged. No less interesting, domain walls propagating faster than the spin wave velocity can spontaneously emit spin waves[5,6] – a magnetic analogy of the Cherenkov Effect.[7] These options are all well encompassed within a bottom-up approach: a controlled growth of ferromagnetic nanowires of (preferably) semiconductor origin, may serve as an ideal route to provide the necessary architecture and resulting functionalities. This is predominantly the reason of the strong ongoing interest in the elaboration of nonequilibrium growth of (Ga,Mn)As dilute ferromagnetic semiconductor (DFS) nanowires (NWs)[8-21] which would exhibit or even surpass the excellent micromagnetic properties of the prototype, epitaxial (Ga,Mn)As layers.[22] This challenge has proven to be a formidable one. Despite numerous reports claiming various magnetic effects, none of them had provided convincing evidence of the existence of a true long-range ferromagnetic ordered state. Either the hole density $p$, a necessary ingredient of the ferromagnetism in dilute ferromagnetic semiconductors[23] was too low,[24] or the Mn distribution was not random enough[25] to result in a thermodynamical phase transition.

(Ga,Mn)As is routinely grown by low temperature molecular beam epitaxy (MBE) on planar, lattice-matched substrates [most often GaAs(100)].[14] If grown as planar epitaxial layers, it crystallizes in the cubic zinc-blende (ZB) structure typical for GaAs and its solid solutions (with Mn, In, or Al). However, as it has been recently demonstrated by us,[20] (Ga,Mn)As can also be crystallized in the hexagonal wurtzite (WZ) structure, if deposited on the side facets of WZ (Ga,In)As NWs. Electronic properties of systems consisting of hexagonal crystals are affected by the presence of built-in electric fields due to the spontaneous polarization.[26] These effects have been experimentally observed in WZ GaAs NWs, but they are much weaker than in the III-Nitrides.[27] This can also affect the magnetic properties of (Ga,Mn)As in the WZ phase. As far as the magnetic properties are concerned, despite some theoretical predictions,[10,21] it is not sure how the hexagonal (wurtzite) structure of (Ga,Mn)As influences the magnetic properties of this material, extensively studied over the last 2 decades in its native cubic ZB phase.[23,24] The first results of calculations of magnetic exchange energies of Mn atoms located in the Ga sites of WZ GaAs are presented in this paper (see the paragraph "Theoretical modelling"). Moreover in the limited number of reports concerning GaAs-(Ga,Mn)As core-shell NW structures[9,15,16] either mixed WZ-ZB phase, or defected (with high density of stacking faults) ZB phase (Ga,Mn)As shells were investigated. In all these reports either the SQUID magnetometry procedures, or sample preparation routines for SQUID specimens didn't allow to unequivocally distinguish between the short range superparamagnetic, and long range ferromagnetic order of magnetic moments of $Mn_{Ga}$ ions in (Ga,Mn)As. In only one report, the magneto-transport properties of a single GaAs-(Ga,Mn)As core shell NW were investigated,[16] for the NW with mixed WZ-ZB phases.

However, even in this case, no typical features known for (Ga,Mn)As layers[24] or (Ga,Mn)As stripes etched from (Ga,Mn)As layers[28,29] pointing on the long-range FM order were evidenced. In our first report on all wurtzite (Ga,In)As-(Ga,Mn)As NWs the hole density was found to be too low to support the FM state.[20]

Two classes of samples comprising GaAs NWs uniformly doped with Mn were studied quite recently: (i) MBE grown uniformly doped GaAs:Mn NWs with very low Mn content (with Mn concentrations at the doping level, i.e. $10^{17} - 10^{18}$ $cm^{-3}$),[13,17] (ii) highly defected GaAs:Mn NWs obtained by Mn ion implantation into MOVPE grown GaAs NWs.[11,12] In both cases no features characteristic for (Ga,Mn)As dilute ferromagnetic semiconductor were found, due to much too low Mn and/or valence band holes content.

In this Letter we report on the magnetic properties of very thoroughly characterized, complex mutli-shell NWs, adequately engineered to reduce the hole leakage out of the Mn-containing layer. In these structures the (Ga,Mn)As shell is typically sandwiched between an inner (Ga,Al)As and outer LT GaAs shells. Both materials are acting as barriers for holes. These NWs display undisputedly and for the first time the existence of a spontaneous magnetic moment – the ultimate evidence that the long range FM order has been realized in a MBE grown, quasi-one-dimensional (Ga,Mn)As structure.

## Experimental results and discussion

Wurtzite (Ga,In)As-(Ga,Al)As-(Ga,Mn)As-LTGaAs core-multishell NWs were grown with Au catalyst on GaAs(111)B substrates. The average surface density of NWs estimated from SEM images is at the level of $10^9$ $cm^{-2}$. The NWs are between 3.5 - 4.0 µm long and their typical diameter increases from about 70 nm at the bottom part (near the substrate) to a maximum of 200 nm at the 3/5 of their height. At the upper ends, just below the gold catalyzer the diameter falls into 120 - 150 nm range. The side-facets of the upper parts of the NWs are rougher than at the lower parts (see Fig. 1).

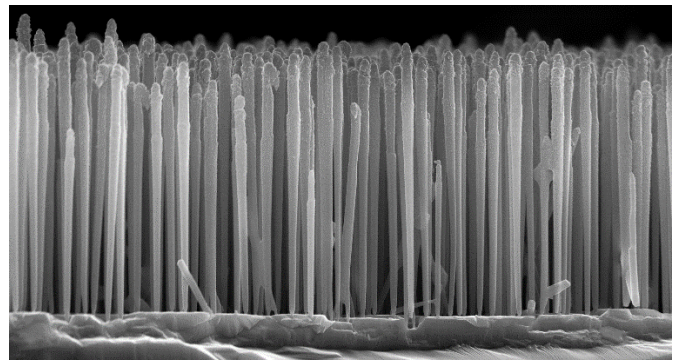

**Fig. 1.** SEM image (cross-sectional view) of (Ga,In)As-(Ga,Al)As-(Ga,Mn)As-LTGaAs core-multishell nanowires grown by MBE on GaAs(111)B substrate with 1Å thick predeposited gold layer.

The typical side walls morphology and the internal structure of the NW are shown in the TEM images displayed in Fig. 2.

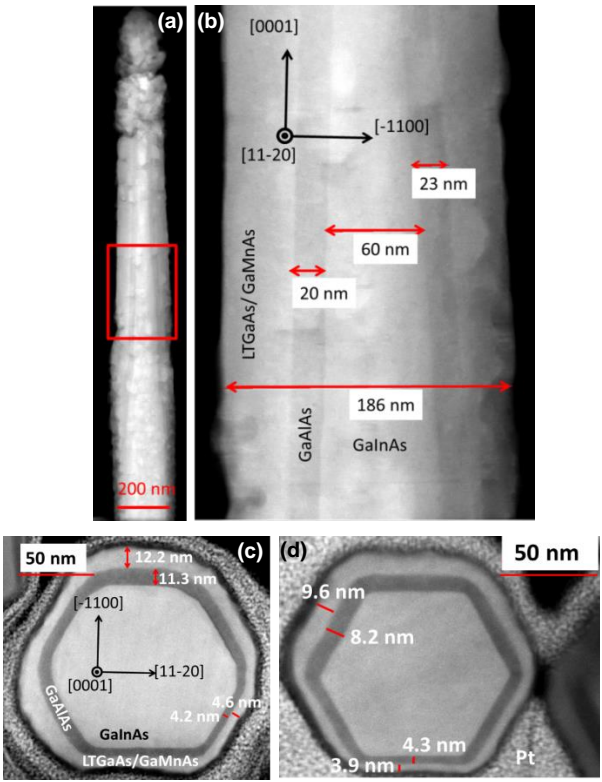

**Fig. 2.** STEM-HAADF images of NWs. (a) top-half part of NW in [11-20] zone axis; (b) zoomed part of (a); (c), (d) - FIB cross-sections of two representative NWs (bottom parts). Technical details concerning these images are included in the Experimental Section

The plan views of a NW placed on a TEM grid with different magnifications are shown in Fig. 2(a) and 2(b). In the bottom part the NW conserves the perfect hexagonal form with well-defined {1-100} facets and sharp corners visible in the cross-section [Fig (2d)]. In the middle and upper part the additional {11-20} facets start to form. The (Ga,Mn)As and LTGaAs shells are grown at low temperature therefore external surfaces of the NW do not conserve well defined corners and become round. The diameter and morphology of the (Ga,In)As core and the thicknesses of the (Ga,Al)As and (Ga,Mn)As shells at the mid- and the top part of the NWs are indicated at cross-sectional images presented in Fig. 2(c) and 2(d), respectively.

It is clearly visible that the NW diameter as well as the thicknesses of the shells is changing significantly along the nanowire length and depend on the orientation of the NW facets with respect to neighboring NWs and incoming flux of elements. This is due to the shadowing effects of NWs with the axes inclined at about 20 deg, with respect to the impinging fluxes emitted by the effusion cells, which are important at this surface NW density. The (Ga,Al)As shell is about 10 nm thick at the bottom part of NW and can reach 20-25 nm in the middle part of the NW. In the top part of nanowires, where the diameter of the cores becomes smaller the low temperature grow rate of Ga(Mn)As and LT-GaAa outer shells is much faster due to the direct exposure to the fluxes of elements, in comparison to the bottom NW parts which are almost totally shadowed.

For this reason the thickness of the (Ga,Mn)As/LT-GaAs shells changes from 5 nm at the bottom part of NW up to 45-50 nm in the widest parts directly exposed to the element fluxes.

The focused ion beam (FIB) cross-sections of the bottom and middle parts of NWs displayed in Fig, 2(c) and 2(d) show that the shells are continuous around NWs even in their thinnest parts. Continuity of the shells has been checked in the bottom part of NW by 3D EDS tomography (see Fig. 3).

Because of the fast grow rate, in the upper part of the NW the LT-GaAs shell becomes rough (RMS $\cong$ 10÷30nm) and thick, in comparison to (Ga,In)As core and (Ga,Al)As shell. This local variation of the thicknesses of outer (LTGaAs and (Ga,Mn)As shells makes it difficult to analyse STEM images obtained from the plan view projection. But we know from investigation of the similar nanowires, (but grown without outermost LT-GaAs shells) that (Ga,Al)As shell is continuous up to the top of the (Ga,In)As core. However the topmost part of the NW consists also partially from axial (Ga,Al)As and Ga(Mn)As grown over Ga(In)As .

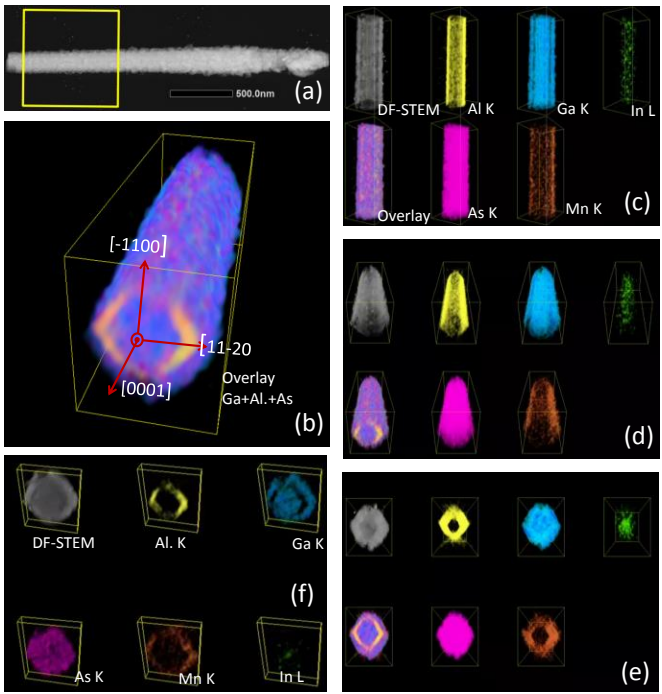
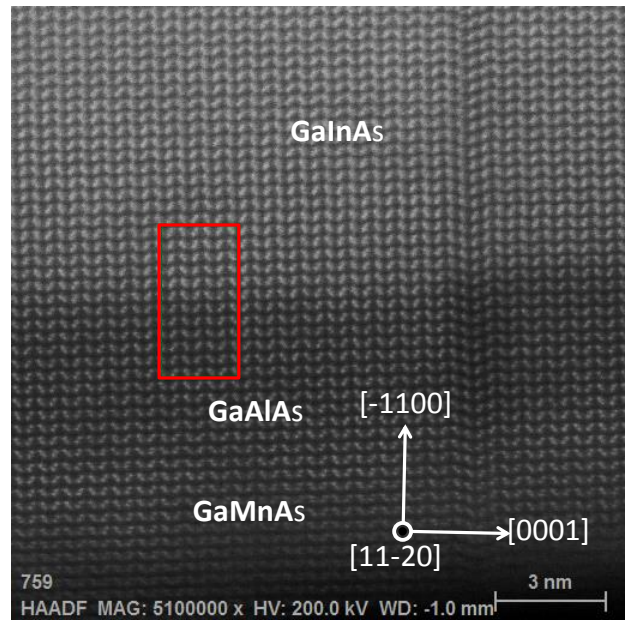

**Fig. 3.** EDS tomography of an individual as-grown NW; (a) - STEM-DF image, yellow frame indicate tomography reconstruction region which is a bottom NW part; (b) - 3D overlayer of As, Al, Ga, Mn, K and In L signals; (c,d,e) - 3D reconstruction in a perspective view of elements spatial distribution seen from different angles (the apparent thicknesses of Al-rich and Mn-rich layers are related to the perspective projection); (f) - slab-cross-section visualization of the elements distribution in the shells.

Figure 3(a) shows the NW region where tomography was performed. In Fig. 3(b) the volumetric renderings of 3D over layer of EDS signals from Al, Ga, In, Mn and As are shown. The upper and bottom parts of reconstruction show loss of the resolution which is due to the limited tilt (±60°) during the tomography procedure. Close inspection of the 3D data proves that (Ga,Al)As and (Ga,Mn)As shells are continuous. The shape of the core can be determined rather from Al signal than In, due to the small concentration of indium. The outer shell containing Mn can clearly be seen in these 3D EDS renderings outside (Ga,Al)As inner shell. The Mn concentration along the NW axis is homogenous. Atomic resolution STEM-HAADF image of the edge of nanowire, together with EDS elemental mapping is shown in Fig. 4.

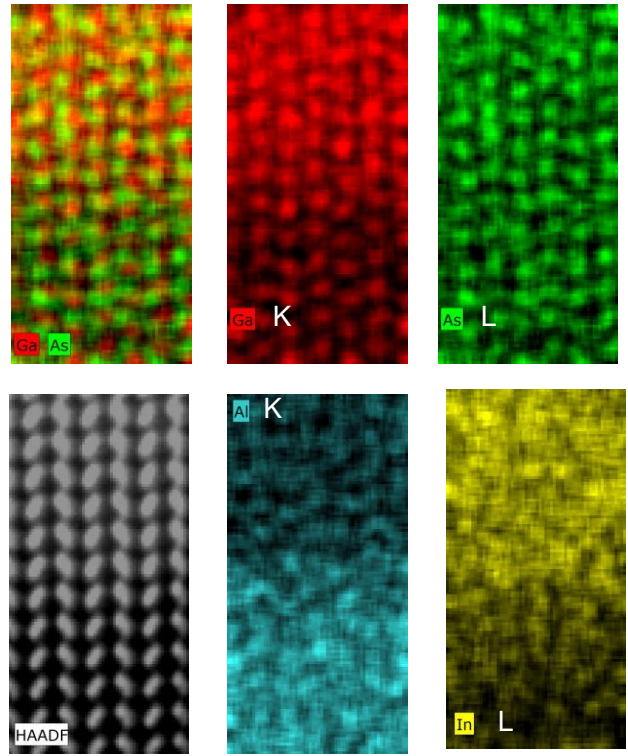

**Fig. 4.** Probe corrected STEM-HAADF image of the edge of the nanowire, and near atomic resolution EDS mapping of (Ga,Mn)As/(Ga,Al)As shell. The STEM-HAADF image obtained during elemental mapping has lower resolution due to the drift.

HR-STEM image confirms that the core and the shell of NW have wurtzite structure. Propagation of the SF defects from the core to the shell proves the epitaxial mechanism of the shell growth. Near atomic resolution EDS mapping of the (Ga,Mn)As/(Ga,Al)As/(Ga,In)As interface region indicates that the core-shell interfaces are abrupt at the atomic scale. The composition of the NW constituents derived from EDS analysis are: 8-10% In in the core, 35%-40% Al in the inner shell, 5% Mn in the (Ga,Mn)As shell. The Mn distribution in (Ga,Mn)As shell is homogenous.

Several NWs have been studied by TEM and in all of them hexagonal (In,Ga)As cores enforced the same WZ structure on the inner (Ga,Al)As, (Ga,Mn)As and the outer LT GaAs shells, all are found in a perfect epitaxial relation to the preceding ones. The SF density is similar to that reported in [Ref. 20], except only the top-most NW part, where the axial grow can still proceed after closing the effusion cells shutters, and can result in formation of 10-20 nm long segments with twinned cubic structure or 4H hexagonal polytype. The average density of SFs is in the range 20÷60 SF/micrometer of the NW length. The In content (about 10 %) in (Ga,In)As cores has been deliberately chosen to set their lattice parameter to be substantially higher than that of (Ga,Al)As and (Ga,Mn)As shells. As can be seen on TEM images shown in Fig. 4 the interfaces between the core, inner (Ga,Al)As and middle (Ga,Mn)As shell are dislocation free, which means that both shells are fully strained to the core. In the core-shell NWs comprising materials with different lattice parameters both cores and shells are subjected to strain[30] (in contrast to epilayers grown on thick, lattice mismatched substrates). In the NWs studied here the (Ga,In)As cores are under compressive strain, whereas all - (Ga,Al)As, (Ga,Mn)As and LT GaAs shells - are under a tensile strain. This kind of strain in (Ga,Mn)As has important consequences for the magnetic anisotropy of this DFS material, as discussed in the paragraph below.

The wurtzite structure and the chemical composition of the nanowires have been further confirmed by room-temperature Raman scattering (RS) measurements, however no information on free carrier density could be inferred from the data despite the fact that the measurements were performed on the same Raman setup and following the same methods described in our previous study.[20] In the present case the typical RS spectra (not shown) consists of AlAs-like, InAs-like, and GaAs-like phonon modes (displaced from their original positions due to various amount of strain these shells are subjected to), which overlap with the expected position of the hole-concentration-induced, coupled plasmon-LO-phonon (CPLOM) mode in (Ga,Mn)As shell, thus hampering the hole density estimation form the RS data.

Magnetic investigation of the NWs were carried out both in near zero and in strong (H -> 7 T) magnetic field domains with the magnetic field applied both perpendicularly and parallel to the NWs, however, as the results reported here obtained in the weak field domain show minor quantitative differences, for the clarity of the presentation, only the results obtained in the perpendicular orientation are shown. All the relevant details of the procedures employed during the magnetic measurements are given in the Methods. The summary of the magnetic studies of the NWs separated from the growth substrate is presented in Fig. 5.

The main result is the direct observation of a *spontaneous* magnetic moment, $m_s$, manifested itself as the non-zero magnetic signal which appears below about 25 K every time the NWs are cooled down at the carefully prepared $H \cong 0$ conditions. The most relevant experimental traces are marked as red triangles in panels (a-c) of Figure 5 and they constitute the main advance over the previously obtained results for the single shell (Ga,In)As/(Ga,Mn)As NWs,[20] for which no (zero) magnetic moment was observed at the same experimental conditions.

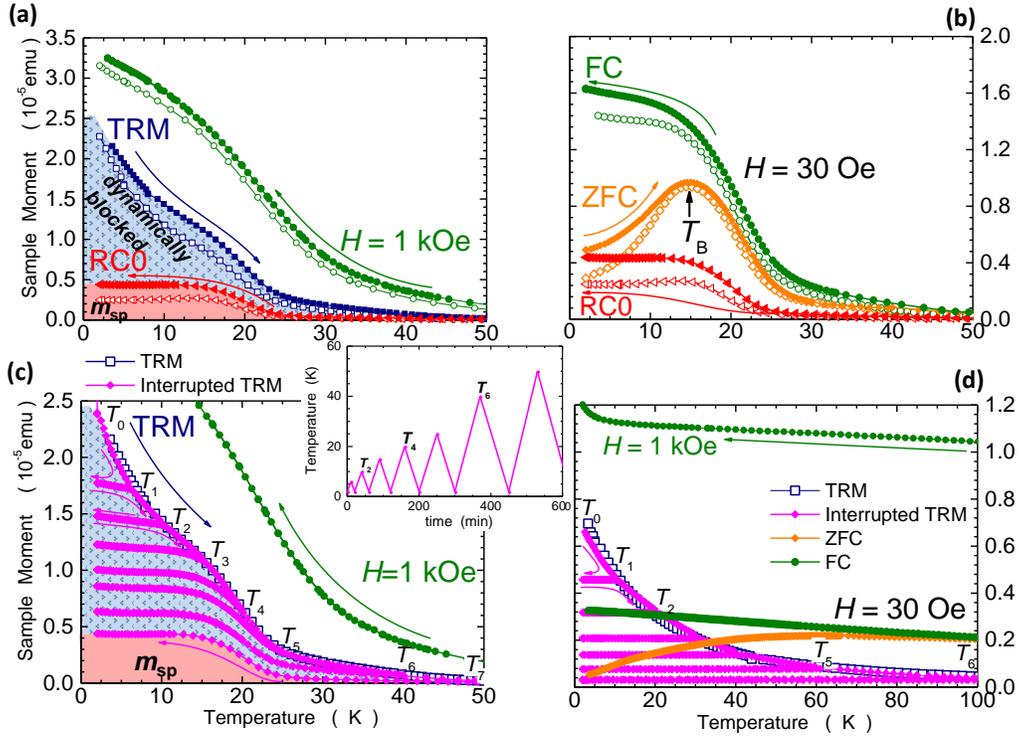

**Fig. 5.** The summary of the weak magnetic field studies of $Ga_{0.90}In_{0.10}As$ /$Ga_{0.65}Al_{0.35}As$/$Ga_{0.95}Mn_{0.05}As$/LTGaAs radially overgrown core/multi-shell nanowires (NWs). The field is applied perpendicular to the NWs.

Open symbols stand for the as grown NWs, full symbols are for the annealed ones: panels (a-c) 4 h at 180 °C at ambient (LTA), panel (d) 30 min at 450 °C inside the MBE chamber. The bluish patterned backgrounds mark the dynamically blocked part of the low field magnetic moment, the reddish one indicates the spontaneous moment ($m_s$) observed on cooling the NWs from (usually) above 50 K at $H \cong 0$, as inferred from panels (a-c).

**(a)** - The field cooled moment at 1 kOe (green, circles), the thermoremanence (TRM, navy squares) and the re-cooling curve at $H \cong 0$ (RC0, red triangles).

**(b)** - Zero field cooled (ZFC) and field cooled (FC) moment measured at 30 Oe after cooling the NWs at $H = 0$ (RC0, triangles). $T_B$ marks the mean blocking temperature of this ensemble of NWs. Please note about 2/3 increase of $m_s$ after LTA as indicated by much larger magnitude of the red triangles marked RC0 cooling trace to that of as grown NWs (open triangles).

**(c)** - Thermal cycling of the NWs at its remanence according to the temperature pattern presented in the inset. This result is superimposed on the typical (continuous) TRM dependence on $T$ (navy squares).

**(d)** – The same types of measurements as described in (b) and (c) but obtained for high temperature annealed NWs with (Ga,Mn)As shell decomposed into MnAs nanocrystals embedded in the GaAs shell.

The importance of this finding stems from the fact that the existence of such a spontaneously appearing magnetic moment is the hallmark of the presence of a long range magnetic order in the investigated material. This means that in the case of the core/multi-shell NWs studied here, we witness an onset of a true, long range ferromagnetic coupling, and that this feature is reported for the first time in (Ga,Mn)As in a quasi- one dimensional form - the wurtzite cylindrical shells in this particular case. We assign this effect to an increased averaged hole density in the (Ga,Mn)As shells owed to the embedment of the (Ga,Mn)As by the double: inner-(Ga,Al)As and outer-LT-GaAs barriers for holes.[31]

It has to be underlined that in general the magnetic results reported here share plenty commonalities with those of the single shell (Ga,In)As/(Ga,Mn)As NWs.[20] Namely, (i) the sizable low-temperature magnetic moment seen during the field cooling at 1 kOe and during recording of the remnant moment on warming (TRM) [Fig. 5(a)]; (ii) well developed anisotropic and slowly saturating, even at strong magnetic fields magnetic hysteresis loops (not shown); (iii) the presence of a clear maximum on the zero-field-cooled (ZFC) moment [Fig. 5(b)]; and (iv) a very similar temperature pattern seen during the thermal cycling of the TRM [Fig. 5(c)]. All these findings indicate that, as already exhaustively discussed in Ref [20], the main magnetic constitution of the NWs has not changed. Magnetically, they can still be regarded as an ensemble of independent mesoscopic volumes (on average $15^3$ nm$^3$ in volume[20]) resulting from the presence of quantum critical fluctuations in the local density of states which are characteristic to the proximity to the metal-insulator transition (MIT). In these volumes the local hole density (LHD) gets sufficiently enlarged to support a *local* ferromagnetic (FM) order due to the Zener mechanism.[32] In the remaining part of the material where the hole density remains too low to support the FM order.[23,33] Therefore, predominantly, characteristics typical for a blocked superparamagnet (BSP) are exerted. The BSP part of the total NWs remnant moment is elaborated during a specially designed thermal cycling of the TRM [detailed in the Methods, and presented in Fig. 5 (c) and marked in bluish pattern in panels (a) and (c) of Fig. 5. However, as already highlighted, the main difference to the previous case is the existence of $m_s$ (marked in light red background in the same panels), which instructs us that a (small) fraction of these mesoscopic volumes got locked into a percolating network along which the magnetic information can be shared over macroscopic distances. Importantly, in the frame of the near MIT LHD fluctuations the establishment of the percolating long range order does not require any sizable increment in $p$. The size of the FM-coupled volumes is related to the hole localization radius which increases strongly as the hole density approaches the critical one.[34] We argue therefore that even for a minute increase of $p$ a statistically relevant number of these volumes might have swelled sufficiently to establish the long range Mn order in a percolation fashion, announced here as a non-zero $m_s$. The highest temperature at which $m_s \neq 0$, around 25 K, we call hereafter the Curie temperature, $T_c$, of the (Ga,Mn)As shell embedded in these core/multi-shell NWs.

The physical process witnessed here is essentially equivalent to the already exploited one in an opposite experiment performed on a thin *ferromagnetic* (Ga,Mn)As layer which had been brought to the verge of hole localization from the metallic side by an external electrical gate.[34] There, the near complete depletion of the layer resulted in a multiple reduction of its spontaneous moment $m_s$. This moment eventually becomes small comparable to that part of the layer's moment which exerted BSP characteristics, which is weakly affected by the reduction of $p$ exactly as it is observed here (cf. Fig. 3(a) and +12 V case in Fig. S3 of Suppl. Inf. to Ref. [34]). It can be said that the relative strength of these two magnetic contributions is set by the ratio of the percolating and independent ferromagnetic volumes, which is an increasing function of $p$. Also, in the authors view not accidentally, the temperature at which $m_s$ sets in both studied here NWs and in the nearly completely depleted thin (Ga,Mn)As layer is very similar – this fact strongly underlines the decisive role of the hole density and gives a strong support for further technological effort to increase it in NWs well beyond the critical region. Quite remarkably, in line with the scenario outlined above, a strong sensitivity to $p$ is observed here. As presented in panels a-c of Fig. 5, while the magnitude of the field cooled moment and the TRM increases only very little, the magnitude of $m_s$ increases by about 2/3 after subjecting the NWs to low temperature annealing at 180 °C, a procedure which is known to increase $p$ in (Ga,Mn)As,[35] but is relatively less effective in capped layer,[36,37] and so the increase in $p$ must have been too small to change the percolating character of the long range ordering what left the magnitude of $Tc$ virtually unchanged.

It should be also noted that, actually, the region of hole concentration so sizably below $p_c$ has never been a subject of an elaborate, near H = 0 magnetic studies in (Ga,Mn)As, so the data accumulated here can serve as a first-hand information on the very low p end of the $p$-$Tc$ magnetic phase diagram. In this context one more interesting finding emerges from our studies. As presented in panels (a-c) of Fig. 5 the BSP part of the NWs' TRM signal extends to temperatures sizably larger than their $Tc$, indicating a presence FM coupled entities characterized by $Tc$ in excess of 50 K. Firstly, we note that our structural characterization decisively precludes existence of any statistically relevant quantity of phase separated FM nanocrystals. We therefore explain the presence of these high-$Tc$ regions in the frame of the critical fluctuations. We argue that this is the proper embedment of the (Ga,Mn)As shell which preserved the hole density in these rare regions in which the critical fluctuations assumed sufficient magnitude to sustain FM order at bulk-(Ga,Mn)As-like temperatures. Indeed, the typically observed $Tc$ in partially compensated (not optimally annealed) 5% (Ga,Mn)As amounts to 50 – 70 K.[22] Although plausible, this line of reasoning can only be substantiated by dedicated magnetic studies in this region of the (Ga,Mn)As magnetic phase diagram. Importantly, the studies should follow the increase of $p$ towards and across the MIT, as the physical constitution of the material like (Ga, Mn)As for $p$ close to $p_c$ may depend whether this particular p has been quenched from "above" or increased from "below".

As it was shown in previous studies frequently (Ga,Mn)As-like NWs growth results in phase separated structure which magnetic properties are set by the presence of FM (typically MnAs) nanocrystals (NCs) in otherwise non-magnetic matrix,[38] or in the coexistence of diluted (Ga,Mn)As and surface segregated MnAs NCs.[8,9,38] In order to exclude a possibility of a magnetic contribution from such NCs in the present study a piece of the NW sample studied here was subjected to a high temperature (HT) annealing at 450 °C for 30 minutes in the MBE growth chamber. This procedure causes decomposition of (Ga,Mn)As solid solution into an ensemble of FM MnAs nanocrystals embedded in the GaAs matrix (see Fig. 6).[38,39]

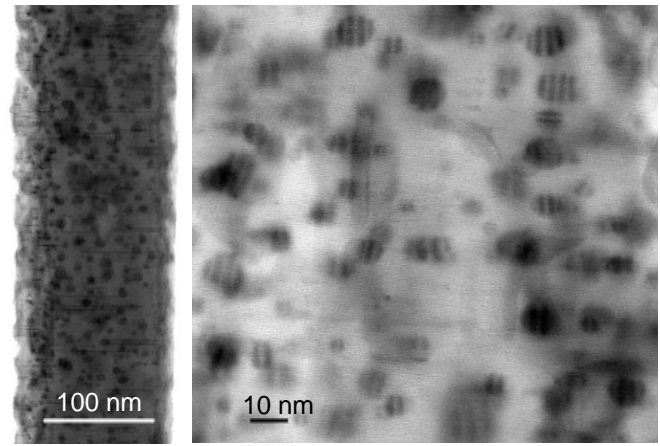

**Fig. 6.** TEM images of a section of NW collected from the piece of sample shown in Figure 1, subjected to high temperature post-growth annealing (at 450 °C for 30 min) causing decomposition of (Ga,Mn)As diluted ferromagnetic semiconductor shell into ensemble of MnAs nanocrystals embedded in GaAs. The alignment of the moiré fringes associated with the nanocrystals points on their preferred orientation with respect to the surrounding GaAs host matrix.

As Fig. 5(d) indicates, indeed, the HT annealed NWs exert BSP characteristics, but differing in every aspect from those of the as-grown NWs. In particular we find that the HT annealed NWs respond to an external field at already elevated temperatures, where any response form the as-grown NWs is hardly seen. This confirms the expected stronger internal FM coupling in MnAs NCs than in the mesoscopic regions with enlarged LHD of the as-grown material. Also, the blocking feature – the maximum on the ZFC curve – is hardly seen in the HT annealed specimen. This further indicates a sizeable average magnitude of the effective magnetic moment of the MnAs NCs NWs, as such broad blocking features are characteristic for systems in which magnetic dipolar interaction between individual NCs (their magnetic moments) starts to dominate their dynamics. Interestingly, these effects are turning particularly strong in reduced dimensionality systems (Ref. [40], and the references therein) i.e. interfaces or NWs, as shown recently in (Ga,Fe)N and (Zn,Co)O layers.[41,42]

Finally, we note that in accordance with the abovementioned studies, in a system containing only separated FM NCs no spontaneous moment is seen. In the panel (d) of Fig. 5 none of the traces recorded during the thermal cycling of the TRM hints to a formation of a spontaneous moment in the whole temperature range.

## Theoretical results

To understand the nature of magnetic exchange in Mn doped WZ GaAs we have carried out a theoretical investigation within the density functional theory (DFT) framework for Mn dopants in bulk WZ GaAs. Since the Mn doped GaAs shell of the NW is rather thick (> 10 nm in the widest part of the NW), these bulk calculations can provide relevant and useful information on the magnetic properties of the system and computationally are considerable less intensive than similar calculations carried out on NW geometries, which are also feasible only for rather small cross sections (< 10 nm). The bulk calculations were performed using a full potential all-electron method with a linearized augmented plane wave + local orbital (LAPW+lo) basis, as implemented in the Wien2k ab-initio package[43], and used the generalized gradient approximation (GGA) with the Perdew-Burke-Ernzerhof (PBE) exchange correlation functional.[44] We constructed bulk supercells of three different sizes containing 72, 108 and 120 atoms using experimental lattice constants.[45] For each case, two Ga atoms are substituted by two Mn atoms. This would correspond to an average doping of 4 %. Figure 7 shows the different arrangements of Mn pairs relative to the crystal axes investigated in this work. The effect of relaxation was not very significant.

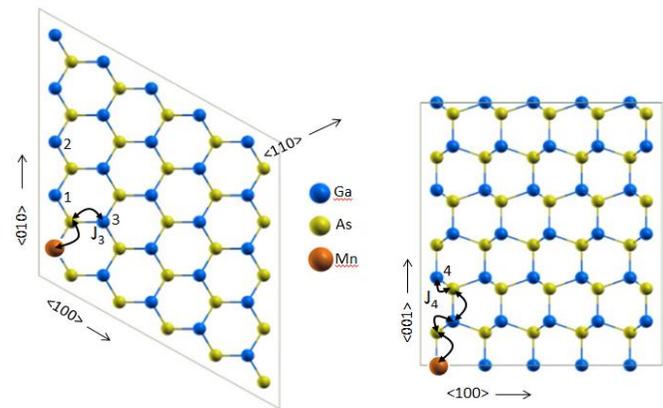

**Fig. 7.** The arrangements of different Mn pairs relative to the different crystal axes in a WZ GaAs nanowire. The labels 1, 2, 3 and 4 at the Ga sites correspond to the position of the 2-nd Mn atom. The distance between the corresponding pairs are d1=d3=3.934, d2=7.978 and d4=6.564 Å, respectively. The calculated exchange energies for corresponding pairs are J1,=216, J2=131, J3=212 and J4=36 meV, respectively. The black curve shows the exchange path between the two sites.

The exchange energy is calculated from the difference in energy when the magnetic moment directions of the Mn pair are aligned parallel and anti-parallel,

$$J = E_{\uparrow\uparrow} - E_{\uparrow\downarrow} \qquad (1)$$

Our calculations show that the ground state is ferromagnetic for all the arrangements of the Mn pair. The exchange is strongest when the Mn pairs are arranged along any crystal-axis direction in the xy-plane, and within the plane the exchange energy is isotropic due to hexagonal symmetry. The exchange is found to be the weakest along the hexagonal axis. We note that while the distance $d_2$ between the Mn pair along <010> direction is about 1.4 Å larger than the distance $d_4$ along <001> direction, the exchange is about four times stronger along <010>. A similar calculation with the ZB structure shows that the exchange energies are comparable to that of WZ except that exchange is strongest along the <110> direction of the ZB structure, and is smaller along the equivalent three cubic axes, which have the same value. The Table 1 shows the comparison of exchange energies for WZ and ZB GaAs.

These calculations show that, concerning the ferromagnetic properties of substitutional Mn dopants in bulk GaAs, the WZ structure is not inferior to the more common ZB structure. Several other effects might play a role in determining the ferromagnetic transition temperature in the multi-shell NWs considered in the experiment. These include primarily the presence of non-substitutional dopants and defects, e.g. interstitial Mn. Preliminary first-principles results indicate that the presence of interstitial Mn dopants favors antiferromagnetic coupling with nearby substitutional Mn, which can significantly influence the overall exchange energy [Md F. Islam, C.M. Canali and J. Sadowski, unpublished]. A systematic study of combined substitutional and interstitial dopants in the WZ structures will be reported in a separate communication.

| Mn placement relative to crystal axis | Wurtzite structure | | | Zinc blende structure | | |
|---|---|---|---|---|---|---|
| | Mn-Mn distance (Å) | Exchange Energy (meV) | Magnetic ground state | Mn-Mn distance (Å) | Exchange Energy (meV) | Magnetic ground state |
| 010 (1 in Fig 5) | 3.9341 | 216 | FM | 5.6446 | 52 | FM |
| 010 (2 in Fig 5) | 7.9780 | 131 | FM | - | | |
| 110 | 3.9351 | 212 | FM | 3.9913 | 211 | FM |
| 001 | 6.5640 | 36 | FM | Equivalent to 010 or 100 | | |

**Tab. 1** Comparison of exchange energies between two Mn atoms in wurtzite and zinc blende structures.

## Methods

### MBE growth:

WZ phase (Ga,In)As core NWs were grown by MBE using Au-catalyzed vapor-liquid-solid (VLS) growth on GaAs(111)B substrates. The $Ga_{0.90}In_{0.10}As$ NW cores were grown at about 500 °C following a short (5-10 min) deposition of pure GaAs at the substrate temperature of about 540 °C. After completing the $Ga_{0.90}In_{0.10}As$ NW cores growth, the substrate temperature was reduced to 400 °C and the first, about 10 nm thick $Ga_{0.65}Al_{0.35}As$ shells were deposited. Afterwards the substrate temperature was further reduced to 230 °C to grow about 10 nm thick $Ga_{0.95}Mn_{0.05}As$ shells at almost stoichiometric ratio of As/(Ga+Mn) impinging fluxes (as calibrated earlier). Finally, the outermost, thin 2 nm shells of low temperature (LT) GaAs have been deposited.

### TEM specimen preparation

TEM investigations were performed on the NWs mechanically transferred to a carbon holey film supported by a standard mesh 300 copper grid from PaciqueGrid, whiteout using any solvent. The cross-section of individual NWs was prepared with FEI Helios 600 nanolab dual beam FIB. The NWs were embedded into platinum from metalorganic source with the use of 15 kV electrons and 30 kV Ga ions for metalorganic compound to selective platinum deposition. Final thinning was performed at 2 kV. Most of TEM investigations were performed in STEM and TEM mode using FEI Titan 80-300 microscope equipped with the image corrector operating at 300 kV.

**TEM imaging and elemental mapping**

Most of TEM investigations were performed in STEM and TEM mode using FEI Titan 80-300 microscope equipped with the image corrector operating at 300 kV. EDS tomography of individual NWs was performed with the use of JEOL JEM-2800 microscope operating at 200 kV equipped with two 100 mm$^2$ SSD detectors. During tomography the half of the standard 3mm grid was mounted in a specially designed holder permitting rotation along alpha axis in the angular range of +/-60°. STEM-HAADF images were taken for a camera length of 91 mm (Figure 2a and 2b) and 76 mm (Figure 2c and 2d) which is suitable to detect the chemical contrast originating from different averaged nuclear numbers Z of Al, and Ga elements. In such conditions lighter (Ga,Al)As appear darker. STEM images taken along the [11-20] zone axis permit to determine clearly the interfaces between different shells along the NW length. The tomography was performed with the following parameters: electron probe size 0.5 nm, tilting angular range +/-60°, degree steps: 2° (STEM) and 4° (EDS), pixel size: 1 nm(STEM) , 2nm (EDS), image size: 512x512 for STEM and 256x256 for EDS. For 3D reconstruction TEMography™ software was used [www.temography.com]. The Al K, Ga K, In L, As K, Mn K lines signals were used to reconstruct 3D distribution of elements. HR-STEM image shown in Figure 3 has been obtained with the probe corrected FEI Themis microscope operating at 200 kV. Near atomic resolution EDS mapping shown in Figure 4 was performed with the Super X detector and FEI chemistem technology. EDS quantification of the elements concentration in the different regions of NW was performed with 20 mm$^2$ Si EDAX detector and FEI Tia software quantification routine. The FIB cross-sections were used for this evaluation, in order to exclude the projection overlapping effects of the shells.

**SQUID measurements**

For magnetic studies the NWs are embedded in PMMA [poly(methyl methacrylate), a popular e-beam lithography resist] and are separated from the substrate at cryogenic temperatures. Then the PMMA flake containing NWs is attached to a thin rectangular piece of Si of beforehand established diamagnetic response (see Ref. 20 for details). Magnetic properties of the NWs are investigated in Quantum Design MPMS XL Superconducting Quantum Interference Device (SQUID) magnetometer equipped with a low field option. The truly near-zero field conditions in the magnetometer ($H \leq 0.1$ Oe, as established by $Dy_2O_3$ paramagnetic test sample) are achieved by degaussing the magnetometer with an oscillating magnetic field of decreasing amplitude. Additionally, a soft quench of the SQUID's superconducting magnet is routinely performed prior to the zero-field studies such as the thermoremnant magnetization (TRM, the measurement of the remnant moment on increasing $T$) and during thermal cycling of the sample brought beforehand to its remanence (mTRM). Each cycle of mTRM consists of warming up the sample to a progressively higher $T$ followed by its re-cooling to the base temperature of 2 K, following the pattern sketched in the inset to Fig 5 (c). The latter measurement allows distinguishing a decaying (with temperature) part of the sample remnant moment [that is the dynamically blocked one by energy barrier(s)] from that related to the spontaneous magnetization in the equilibrium state in the zero field conditions, if of course such a contribution exists in the specimen. All these magnetic measurements are carried out by strictly observing the adequate for minute signals experimental code[46] and follow the recipes outlined in our recent study.[20]

## Conclusions

In conclusion, we have observed for the first time a ferromagnetic phase transition in wurtzite (Ga,Mn)As NWs - a crystal phase of the canonical dilute ferromagnetic semiconductor which has not been thoroughly investigated so far. In principle, this result opens the road towards technologically viable one-dimensional ferromagnetic cylindrical structures made of a semiconducting material obtained in a bottom up approach. The FM phase transition temperature in WZ (Ga,Mn)As is still substantially lower than that of this DFS in its native zinc-blende phase in thin film geometries. It can be assumed that this is due to the lower concentration of valence band holes and/or enhanced distribution of Mn into interstitial sites in the WZ (Ga,Mn)As lattice. The occurrence of a long-range ferromagnetic order in the WZ (Ga,Mn)As NW shells has been observed only in the purposely designed core/multi-shell NW structures in which (Ga,Mn)As was embedded by an inner and outer shell acting as the barriers for holes [(Ga,Al)As and LT GaAs, respectively]. While the theoretical calculations presented here show that the strength of the ferromagnetic exchange coupling for Mn pairs is comparable in the ZB and WZ structures, these experimental findings call for a systematic theoretical investigation of the magnetic properties of (Ga,Mn)As NWs in the WZ structure. These studies together with advances in the engineering of the NW multi-shell heterostructures may help elucidate the outstanding issue of the low ferromagnetic transition temperature in these dilute magnetic semiconductor nanostructures.

## Acknowledgements

This work has been supported by the research projects Nos: 2014/13/B/ST3/04489 2012/07/B/ST5/02080, and DEC-2012/06/A/ST3/00247 financed through the National Science Centre (Poland). The authors acknowledge Dr Aoyamy Yoshitaka from JEOL Ltd., Japan for performing EDS tomography, Dr Sergei Lopatin from FEI Electron Optics, Eindhoven, Netherlands for help in investigation with the use of probe corrected FEI Titan Themis microscope with Chemistem technology, Msc Bogusława Kurowska from Institute of Physics PAS in Warsaw for preparation of FIB cross-sections. The most part of TEM investigation was performed using equipment founded by European Regional Development Fund through the Innovative Economy grant No. POIG.02.01-00-14-032/08.